# Market-based Microgrid Optimal Scheduling

Sina Parhizi, Amin Khodaei
Department of Electrical and Computer Engineering
University of Denver
Denver, CO, USA
sina.parhizi@du.edu, amin.khodaei@du.edu

*Abstract*— **This paper presents an optimal scheduling model for a microgrid participating in the electricity distribution market in interaction with the Distribution Market Operator (DMO). The DMO is a concept proposed here, which administers the established electricity market in the distribution level, i.e., similar to the role of Independent System Operator (ISO) in the wholesale electricity market, sets electricity prices, determines the amounts of the power exchange between market participators, and interacts with the ISO. Considering a predetermined main grid power transfer to the microgrid, the microgrid scheduling problem will aim at balancing the power supply and demand while taking financial objectives into account. A stochastic programming method is employed to model prevailing uncertainties in the microgrid grid-connected and islanded operations. Numerical simulations exhibit the application and the effectiveness of the proposed market-based microgrid scheduling model.**

## NOMENCLATURE

***Sets and Indices:***

| | |
|---|---|
| ch | Superscript for energy storage charging mode. |
| $d$ | Index for loads. |
| dch | Superscript for energy storage discharging mode. |
| D | Set of adjustable loads. |
| G | Set of dispatchable units. |
| $i$ | Index for DERs. |
| $s$ | Index for scenarios. |
| S | Set of energy storage systems. |
| $t$ | Index for hours. |
| $\tau$ | Index for sub-periods. |

***Parameters:***

| | |
|---|---|
| $c$ | Penalty for scheduled power violation. |
| $DR$ | Ramp down rate. |
| $DT$ | Minimum down time. |
| $F$ | Operation cost function of dispatchable unit. |
| $MC$ | Minimum charging time. |
| $MD$ | Minimum discharging time. |
| $MU$ | Minimum operating time. |
| $pr$ | Probability of scenarios. |
| $P^{sch}$ | Main grid power transfer assigned to the microgrid from the DMO. |
| $U$ | Islanding binary indicator (1 when grid-connected, 0 when islanded). |
| $UR$ | Ramp up rate. |
| $UT$ | Minimum up time. |
| $\alpha, \beta$ | Specified start and end times of adjustable loads. |

***Variables:***

| | |
|---|---|
| $C$ | Energy storage available (stored) energy. |
| $D$ | Load demand. |
| $I$ | Commitment state of dispatchable unit (1 when committed, 0 otherwise). |
| $LS$ | Load curtailment. |
| $P$ | DER output power. |
| $P_M$ | Main grid power transfer determined via optimal scheduling. |
| $\Delta P_M$ | Main grid power transfer mismatch with respect to the assigned value. |
| $\Delta P_M^{+}$ | Positive main grid power transfer mismatch. |
| $T^{ch}$ | Number of successive charging hours. |
| $T^{dch}$ | Number of successive discharging hours. |
| $T^{on}$ | Number of successive ON hours. |
| $T^{off}$ | Number of successive OFF hours. |
| $v$ | Energy storage charging state (1 when charging, 0 otherwise). |
| $u$ | Energy storage discharging state (1 when discharging, 0 otherwise). |
| $z$ | Adjustable load state (1 when operating, 0 otherwise) |
| $\delta$ | Power transfer violation indicator (1 when violated, 0 otherwise). |
| $\upsilon$ | Value of lost load. |

## I. INTRODUCTION

Microgrids provide significant opportunities and new functionalities to the smart electric grid by facilitating the integration of distributed energy resources (DERs) to distribution grids and further increasing system reliability and resiliency. Providing local intelligence to the system, reduction in greenhouse gas emissions, and reducing the need for expanding transmission and distribution facilities as a result of generation-load proximity are among other value propositions of microgrids [1]. In addition, the microgrid capability to be operated as a single controllable entity enables an active participation in variety of demand response programs as well as sell back of electricity to the utility grid at high price hours [2]–[9]. Majority of current microgrid installations has been undertaken in North America [10], but it is expected that by 2020 microgrid deployments be more uniformly distributed around the world [11]. The microgrid market in the U.S. is projected to be an annual $2 billion by 2015. The total installed microgrid capacity is expected to grow from 1.1 GW in 2012 to 4.7 GW with a market opportunity of U.S. $17.3 billion in 2017 [12]. Microgrid deployments are also federally supported in the U.S. The U.S. Department of Energy is planning microgrid developments capable of reducing outage times by more than 98%, reducing emissions by more than 20%, and improving system energy efficiency by more than 20% by 2020 [13].

The microgrid control, one of the most imperative aspects when deploying microgrids, is commonly performed in three hierarchical levels, including primary, secondary, and tertiary [14]. The first two control levels deal with droop control and frequency/voltage adjustment and restoration when there is a change in the amount of microgrid load and/or generation as well as islanding transitions. The third level, however, schedules microgrid components to obtain an economic dispatch of available resources while taking main grid interactions into account. Microgrid scheduling problem aims to minimize the operational costs of local DERs, as well as the energy exchange with the main grid, to supply forecasted load demand in a certain period of time (typically one day). There are two common designs for the control architecture of the microgrid controller: centralized and decentralized. In the decentralized architecture, each component acts as an agent with ability of decision making and communication with other agents [15]. The decentralized architecture makes it easier to expand the scale of the microgrid and is more immune to failures of its components. In the centralized architecture, on the other hand, scheduling will be performed centrally in a central computing unit, which is able to access microgrid-wide generation and load information and dispatch generation according to total load demand and individual generator's cost curves [16]. There are benefits and disadvantages associated with both architectures as discussed in [17]. This paper adopts a centralized architecture for microgrid scheduling as it does not require new investment to build the communication infrastructure and facilitates application of optimization methods for ensuring solution optimality.

A variety of approaches are proposed in the literature to solve the microgrid optimal scheduling problem, including deterministic, heuristic, and stochastic methods. Mixed integer programming (MIP), is widely used to formulate microgrid scheduling problems. Several studies also consider prevailing uncertainties in the microgrid optimal scheduling process, using stochastic programming [18]–[20], chance-constrained programming [21], and robust optimization [15].

Increasing demand-side elasticity and active participation of loads in the power system in response to electricity price variations is highly stressed to operate the system more efficiently and to avoid high price spikes caused by inelastic loads [22]. Microgrids allow an efficient integration and control of large penetration responsive loads which would further increase the demand-side elasticity. Moreover, distributed generators and energy storage enable a highly fast and controllable load. Currently, however, these resources are scheduled based on a price-based scheme; i.e. the microgrid controller determines the least-cost schedule of available DERs and loads, as well as the main grid power transfer, based on the day-ahead market price (which is forecasted by the microgrid or the electric utility). Under this scheme, the utility forecasts an estimate of the microgrids' loads in its service territory and submits to the wholesale market via available mechanisms. Once the electricity price is determined, through the wholesale market, the utility sends the actual prices to microgrids. Although it might seem efficient, this approach has the potential to cause several drawbacks when the microgrid penetration in distribution network is high, including shifting the peak hours. This approach is prone to cause new peaks as there is a high probability that microgrids follow a different schedule as the one forecasted by the utility once actual prices are received, as the demand in responsive loads is inversely proportional to electricity prices. The increase in the number of entities with responsive loads operated by price-based methods would intensify this issue. In other words, setting the price centrally by the utility and sending it to microgrids, so they can accordingly schedule their resources, can potentially result in significant uncertainty in system load profile.

The aforementioned drawback, combined with the enhanced complexity in managing a large number of microgrids in a foreseeable future, make the case for a new approach to the system operation and utility ratemaking in presence of microgrids. In this paper, a market-based microgrid optimal scheduling model is proposed to address the aforementioned problem and increase microgrid-integrated distribution system efficiency. The proposed model is based on a Distribution Market Operator (DMO) model. DMO is an entity which is hosted in the distribution network to manage microgrids interaction with the main grid. Similar concepts as DMO can be found in [23][24], where the transformation of existing utility operations to integrate high penetration microgrids are discussed by introducing Distributed System Platform Provider (DSPP) and in [25] where a price-based simultaneous operation of microgrids and the Distribution Network Operator (DNO) is proposed. In this paper, the microgrid operation under a distribution market is investigated. A microgrid scheduling model is proposed that seeks to optimally schedule the microgrid components while complying with main grid power transfer schedule imposed by the market. Future work will discuss the process by which the DMO schedules the power transfers and operates the distribution market.

The rest of the paper is organized as follows: Section II outlines the model for retail market in the distribution network, Section III presents the formulation for the market-based microgrid optimal scheduling problem, Section IV presents the numerical results and discussions, and Section V concludes the paper.

## II. Distribution Network Market Model

A Distribution Market Operator (DMO) is proposed to cooperate with the Independent System Operator (ISO) in the wholesale electricity market and facilitate establishing a competitive electricity market in the distribution network level to exchange energy and grid services. DMO is a smart platform that enables market activities and grid operations, coordinates with the utility to improve the investment planning, and interacts with the ISO to negotiate awarded demand bids or power purchases for ensuring an efficient system operation. DMO would further facilitate a rapid and widespread integration of microgrids from a system operator's perspective by addressing prevailing integration challenges.

In order to address the problems of price-based microgrid scheduling, a market-based solution to the microgrid scheduling problem is proposed where the microgrid demand is set by the DMO and known with certainty on a day-ahead basis. This will lead to lower peak demands in the system and increased reliability and efficiency of its operation. Establishment of DMO would also be beneficial to the ISO as it allows a significant reduction in the need to invest and expand the communication infrastructure among microgrids and the ISO. DMO can be formed as a new entity or be part of the currently existing electric utilities. An independent DMO would be able to set up a universal market environment instead of one for each utility. It would also be less suspected of exercising market power. On the other hand, a utility-affiliated DMO would be able to perform several functionalities currently possessed by electric utilities without necessitating additional investments.

The microgrid can exchange power with the main grid and act as a player in the electricity market. DMO would serve as an interface between the ISO and microgrids that facilitates microgrids market participation and coordinates the microgrids with the main grid to minimize the risks posed by microgrid operational uncertainties. DMO will receive demand bids from the microgrids, combine them, and offer an aggregated bid to the ISO. ISO will receive aggregated demand bids from DMOs and generation bids from generation companies (GENCOs) to determine the market price with the objective of maximizing the system social welfare. Fig. 1 depicts the interactions of DMO with different players in the market. ISO performs the market clearing process and determines the schedule of each directly connected player. DMO will receive the day-ahead schedule from the ISO and subsequently determine microgrids shares from the awarded power. The microgrid power could be positive (representing the microgrid as a load), or negative (representing the microgrid as a generation). The main grid power transfer to the microgrid would be the amount of power announced to the microgrid by the DMO, hence it would be known to the system operator and therefore eliminate the uncertainties caused by microgrid price-responsiveness. Note that this paper only focuses on modeling the microgrid behavior under distribution market environment, while the detailed DMO will be performed in a future work. The ISO model with responsive loads can be found in [26].

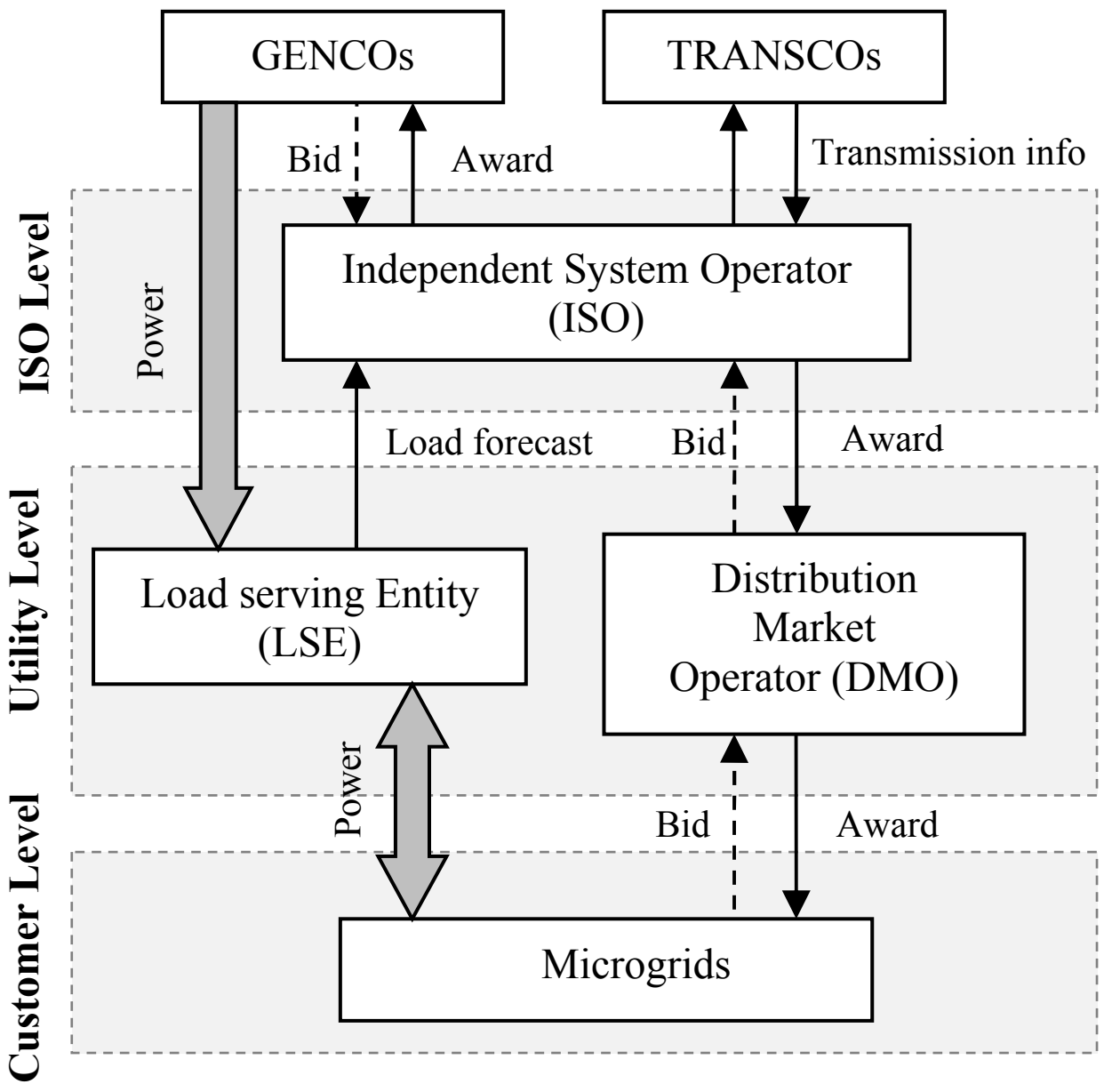

Fig.1 Proposed microgrid market participation through distribution market operator

Implementation of DMO would fix the aforementioned problems that utilities face when they integrate microgrids, but in order for the proposed system to reliably work it is necessary that the microgrid controller schedules its resources based on the scheduled main grid power transfer. Moreover, microgrid controller requires to take microgrid economy into account and find the least-cost schedule while conforming to the scheduled main grid power. This problem is discussed in this paper while considering microgrid uncertainties associated with nondispatchable generation and loads. Several methods, such as robust optimization and stochastic optimization, are used to solve optimization problems with data uncertainty. Robust optimization is a rather conservative approach applied to cases where the solution is desired to be immune against the worst possible uncertain outcome, and/or the information about the probability distribution for the uncertainty is not available. Stochastic optimization is used to handle the optimization problems with random uncertainties where the distribution of uncertainties is known or can be assumed with acceptable accuracy. In a stochastic optimization model, the objective function and constraints depend on not only the optimization variables but random variables with known distribution. The optimization would seek to minimize the expected value of the objective function while satisfying all the constraints for each generated scenario. In this work a stochastic optimization approach is employed to incorporate uncertainties in the scheduling model as the distribution of uncertainties, which are the forecasts in generation of renewable energy resources and loads, is subject to some known distributions obtained from historical data. A large number of scenarios for uncertain parameters is generated and further reduced using available reduction techniques to ensure the computational efficiency of the proposed model.

A 24-h scheduling horizon is considered with sub-periods of 10 minutes for accurately considering nondispatchable generation and load variations. Dispatchable units, energy storage, and adjustable loads are committed hourly but their associated dispatch is performed on a sub-hourly basis. The microgrid submits its demand bid to the DMO one day ahead and the DMO determines the market clearing price and the accepted demand for the entire scheduling horizon. Based on the accepted and known day-ahead demand profile, the microgrid master controller will solve the optimal scheduling problem. It will be assumed that violations from the scheduled value will be penalized based on a pre-established price (which could be simply considered equal to the market clearing price at the microgrid point of common coupling). The penalty, however, will be applied when the violation is positive, i.e., the main grid power transfer is larger than the scheduled power by the DMO, or in other words, when the microgrid appears as a larger load and requests a larger supply of power from the main grid. Negative violation will not be penalized as the microgrid helps with generation in the distribution network by reducing its load.

## III. Market-Based Scheduling Problem Formulation

### A. Objective

The objective of this problem is to determine the least-cost day-ahead schedule of loads, dispatchable generation units, and energy storage in the microgrid (1) when the profile of the main grid power transfer over the scheduling horizon is known (i.e., determined and announced by the DMO). In order to take the associated uncertainties into consideration, a stochastic scenario-based optimization model is employed as proposed in [27]. Each scenario simulates an outcome with a uniformly distributed random nondispatchable unit generation and load.

$$\min \mathrm{E}\left[\sum_{s} pr_s \sum_{t}\sum_{\tau}\left(\sum_{i\in \mathrm{G}} F_i(P_{it\tau s}, I_{it}) + \upsilon LS_{t\tau s} + c\Delta P^{+}_{M,t\tau s}\right)\right] \quad (1)$$

The objective includes three terms of operation cost of dispatchable units (which includes generation cost, and startup/shut down costs), the load curtailment cost (defined as the value of lost load times the amount of load curtailment) and the penalty for deviation from the scheduled main grid power transfer. The objective is weighted, using probability, and summed over all scenarios.

### B. Operational Constraints

The objective is subject to the following operational constraints:

$$\sum_i P_{it\tau s} + P_{M,t\tau s} + LS_{t\tau s} = \sum_d D_{dt\tau s} \qquad \forall t, \forall \tau, \forall s \quad (2)$$

$$P_i^{\min} I_{it} \le P_{it\tau s} \le P_i^{\max} I_{it} \qquad \forall i\in \mathrm{G}, \forall t, \forall \tau, \forall s \quad (3)$$

$$P_{it\tau s} - P_{it(\tau-1)s} \le UR_i \qquad \forall i\in \mathrm{G}, \forall t, \forall \tau, \forall s \quad (4)$$

$$P_{it(\tau-1)s} - P_{it\tau s} \le DR_i \qquad \forall i\in \mathrm{G}, \forall t, \forall \tau, \forall s \quad (5)$$

$$T_{it}^{\mathrm{on}} \ge UT_i (I_{it} - I_{i(t-1)}) \qquad \forall i\in \mathrm{G}, \forall t \quad (6)$$

$$T_{it}^{\mathrm{off}} \ge DT_i (I_{i(t-1)} - I_{it}) \qquad \forall i\in \mathrm{G}, \forall t \quad (7)$$

$$P_{it\tau s} \le P_i^{\mathrm{dch,max}} u_{it} - P_i^{\mathrm{ch,min}} v_{it} \qquad \forall i\in \mathrm{S}, \forall t, \forall \tau, \forall s \quad (8)$$

$$P_{it\tau s} \ge P_i^{\mathrm{dch,min}} u_{it} - P_i^{\mathrm{ch,max}} v_{it} \qquad \forall i\in \mathrm{S}, \forall t, \forall \tau, \forall s \quad (9)$$

$$u_{it} + v_{it} \le 1 \qquad \forall i\in \mathrm{S}, \forall t \quad (10)$$

$$C_{it\tau s} = C_{it(\tau-1)s} - P_{it\tau s}\Delta\tau \qquad \forall i\in \mathrm{S}, \forall t, \forall \tau, \forall s \quad (11)$$

$$0 \le C_{it\tau s} \le C_i^{\max} \qquad \forall i\in \mathrm{S}, \forall t, \forall \tau, \forall s \quad (12)$$

$$T_{it}^{\mathrm{ch}} \ge MC_i (u_{it} - u_{i(t-1)}) \qquad \forall i\in \mathrm{S}, \forall t \quad (13)$$

$$T_{it}^{\mathrm{dch}} \ge MD_i (v_{it} - v_{i(t-1)}) \qquad \forall i\in \mathrm{S}, \forall t \quad (14)$$

$$D_d^{\min} z_{dt} \le D_{dt\tau s} \le D_d^{\max} z_{dt} \qquad \forall d\in \mathrm{D}, \forall t, \forall \tau, \forall s \quad (15)$$

$$\sum_{t\in[\alpha_d, \beta_d]} D_{dt\tau s} = E_d \qquad \forall d\in \mathrm{D}, \forall \tau, \forall s \quad (16)$$

$$T_{dt}^{\mathrm{on}} \ge MU_d (z_{dt} - z_{d(t-1)}) \qquad \forall i\in \mathrm{D}, \forall t \quad (17)$$

The power balance constraint is considered in (2) to make sure that the sum of the main grid power transfer plus the locally generated microgrid power matches the total load, while load curtailment variable is added to ensure that this balance is satisfied at all times. The nondispatchable generation and fixed load values are forecasted in this constraint while they will change in each scenario. Dispatchable unit constraints include generation minimum/maximum limits (3), ramp up/down limits (4)-(5), and minimum up/down time limits (6)-(7). Energy storage constraints include maximum charging and discharging constraints (8)-(9), charging/discharging mode (10), available stored energy limits (11)-(12), and minimum charge/discharge time (13)-(14). Adjustable loads constraints include rated power limit (15), required energy consumption in a certain period specified by [$\alpha_d$, $\beta_d$] (16), and minimum operating time (17). In the case of inter-temporal constraints, such as minimum up/down times, it must be ensured that at the first period $\tau$ of each hour $t$, the constraint holds with respect to the last period of the previous hour.

### C. Main Grid Power Transfer Deviation Modeling

The main grid power transfer for each microgrid is scheduled and assigned by the DMO. However, microgrid can deviate from the scheduled power transfer and pay a penalty as proposed in (1).

$$-P_M^{\max} U_{t\tau s} \le P_{M,t\tau s} \le P_M^{\max} U_{t\tau s} \qquad \forall t, \forall \tau, \forall s \quad (18)$$

$$\Delta P_{M,t\tau s} = P_{M,t\tau s} - P_{M,t}^{sch} \qquad \forall t, \forall \tau, \forall s \quad (19)$$

$$-P_M^{\max} \delta_{ts} \le \Delta P^{+}_{M,t\tau s} \le P_M^{\max} \delta_{ts} \qquad \forall t, \forall \tau, \forall s \quad (20)$$

$$-P_M^{\max}(1-\delta_{ts}) \le \Delta P_{M,t\tau s} - \Delta P^{+}_{M,t\tau s} \le P_M^{\max}(1-\delta_{ts}) \qquad \forall t, \forall \tau, \forall s \quad (21)$$

To model the islanded operation, binary variable $U_{t\tau s}$ is generated in each scenario to model islanding incidents by zeroing out the main grid power transfer (18). The main grid power transfer mismatch from the amount scheduled by the DMO is set by (19). If the main grid power transfer mismatch is positive, the objective is penalized, where $\delta$=1 and $\Delta P^{+}$=$\Delta P$ using (20) and (21).

## IV. NUMERICAL SIMULATIONS

A microgrid with four dispatchable generation units, a nondispatchable unit, five adjustable loads and one energy storage is considered for simulating the proposed market-based microgrid scheduling model. The microgrid characteristics, as well as forecasted values for fixed load and nondispatchable generation, are borrowed from [6]. Table I shows the scheduled main grid power transfer, with a mismatch penalty of $150/MWh.

TABLE I
MAIN GRID POWER TRANSFER SCHEDULED BY THE DMO

| Time (h) | 1 | 2 | 3 | 4 | 5 | 6 |
|---|---|---|---|---|---|---|
| Power(MW) | 0.70 | 5.60 | 4.90 | 5.60 | 6.30 | 5.60 |
| Time (h) | 7 | 8 | 9 | 10 | 11 | 12 |
| Power (MW) | 4.90 | 5.60 | 6.30 | 4.90 | 5.60 | 5.60 |
| Time (h) | 13 | 14 | 15 | 16 | 17 | 18 |
| Power (MW) | 6.30 | 5.60 | 6.30 | 7.00 | 8.40 | 9.80 |
| Time (h) | 19 | 20 | 21 | 22 | 23 | 24 |
| Power (MW) | 11.20 | 10.50 | 9.80 | 7.70 | 6.30 | 5.60 |

A total of 100 scenarios are generated to simulate errors in the forecasted subhourly (10-minute) nondispatchable generation and islanding from the main grid. The optimal commitment of the dispatchable units for all scenarios and the schedules of energy storage, loads, and generation dispatch for each scenario are also obtained. The operating cost is obtained as $39,566. The resulting commitment schedule is given in Table II where bold numbers show the change compared to the case without islanding. During peak times 14-21 all units are committed to supply local loads and also ensure availability of sufficient generation during transition to islanding. The energy storage is also discharged at its maximum power during islanding periods to contribute to the load balance.

TABLE II
DER SCHEDULE CONSIDERING 1-HOUR ISLANDING

| Unit | Hours (1-24) | | | | | | | | | | | | | | | | | | | | | | |
|---|---|---|---|---|---|---|---|---|---|---|---|---|---|---|---|---|---|---|---|---|---|---|---|
| G1 | 1 | 1 | 1 | 1 | 1 | 1 | 1 | 1 | 1 | 1 | 1 | 1 | 1 | 1 | 1 | 1 | 1 | 1 | 1 | 1 | 1 | 1 | 1 | 1 |
| G2 | 1 | 1 | 1 | 1 | 1 | 1 | 1 | 1 | 1 | 1 | 1 | 1 | 1 | 1 | 1 | 1 | 1 | 1 | 1 | 1 | 1 | 1 | **1** | **1** |
| G3 | 1 | **1** | **1** | **1** | **1** | 0 | 0 | **1** | **1** | **1** | **1** | **1** | **1** | **1** | **1** | **1** | **1** | **1** | **1** | **1** | **1** | **1** | 0 | 0 |
| G4 | 0 | 0 | 0 | 0 | 0 | 0 | 0 | 0 | 0 | 0 | 0 | 0 | 0 | **1** | **1** | **1** | **1** | **1** | **1** | **1** | **1** | 0 | 0 | 0 |

The problem is further solved for a variety of islanding hours to show the impact of the number of islanding hours on the microgrid scheduling results. The operation cost for all cases is shown in Fig. 2. The load curtailment is added to the objective as a penalty term with a cost of $10,000/MWh. As the duration of islanding increases, a larger portion of loads needs to be curtailed. This factor together with increasing need for generation of dispatchable units increases the total operation cost.

Sensitivity to the scheduled main grid power transfer is analyzed by considering a reduced 100 scenarios. Fig. 3 shows the operation cost when the main grid power transfer is a fraction of values in Table I. Increasing main grid power transfer would lessen the need to microgrid generation and may accordingly cause some dispatchable units that were already committed to be turned off at some hours, thus it would decrease the operation cost. In case an islanding happens at these hours, since there are fewer units committed and available to generate, there might be more load curtailment. When the power transfer is relatively low, the number of units that turn off is relatively few and even if they turn off at some hours the overall reduction in the generation cost outweighs the increase in load curtailment cost at some scenarios with islanding at those hours. For example, when the power transfer is increased from 0.35 times of values in Table I to 0.5, G3 is turned off at hour 7, load curtailment cost increases from $992 to $1,814 and generation cost reduces from $66,560 to $60,362.

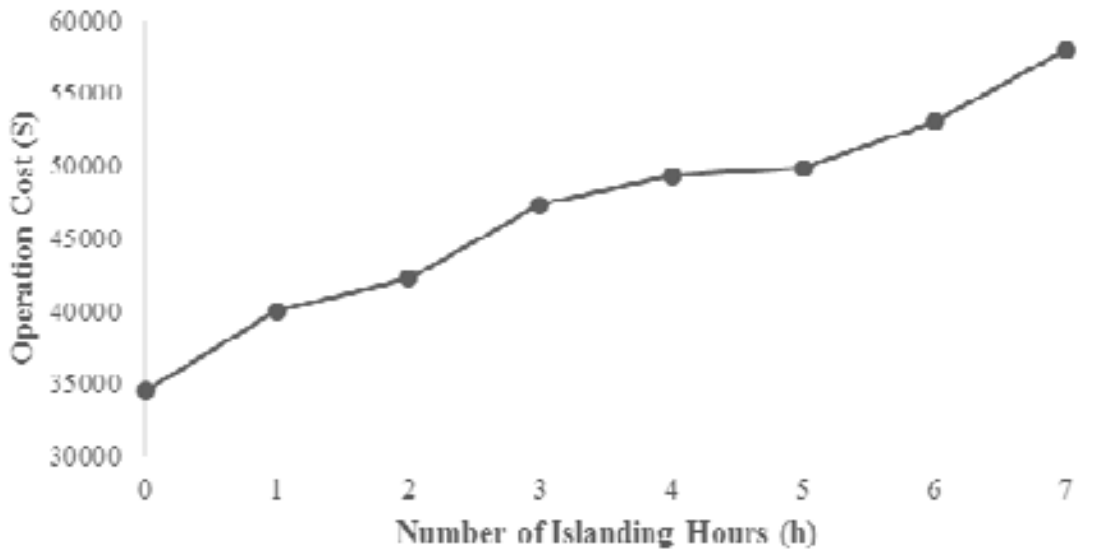

Fig. 2 Microgrid operation cost as a function of number of islanding hours

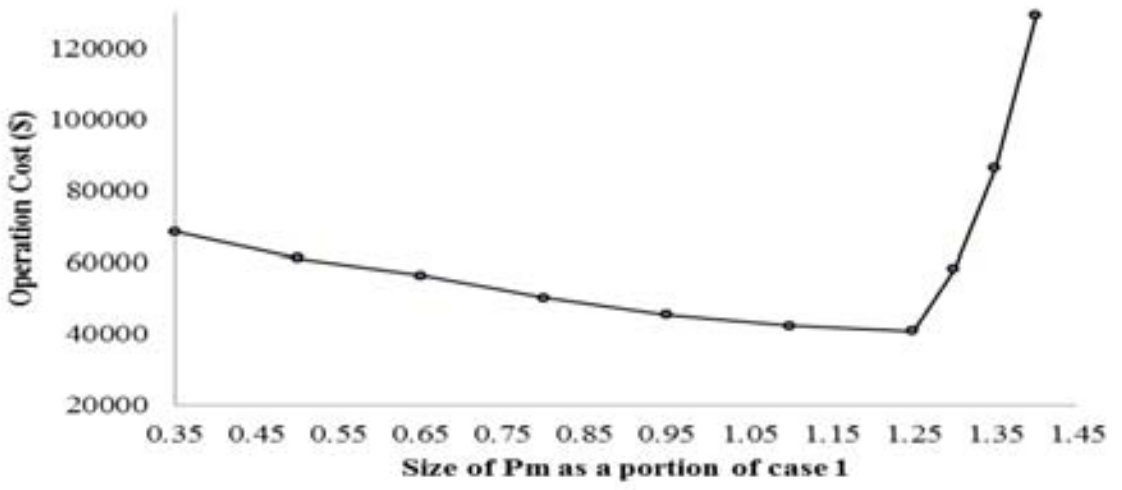

Fig. 3 Microgrid operation cost as a function of main grid power transfer

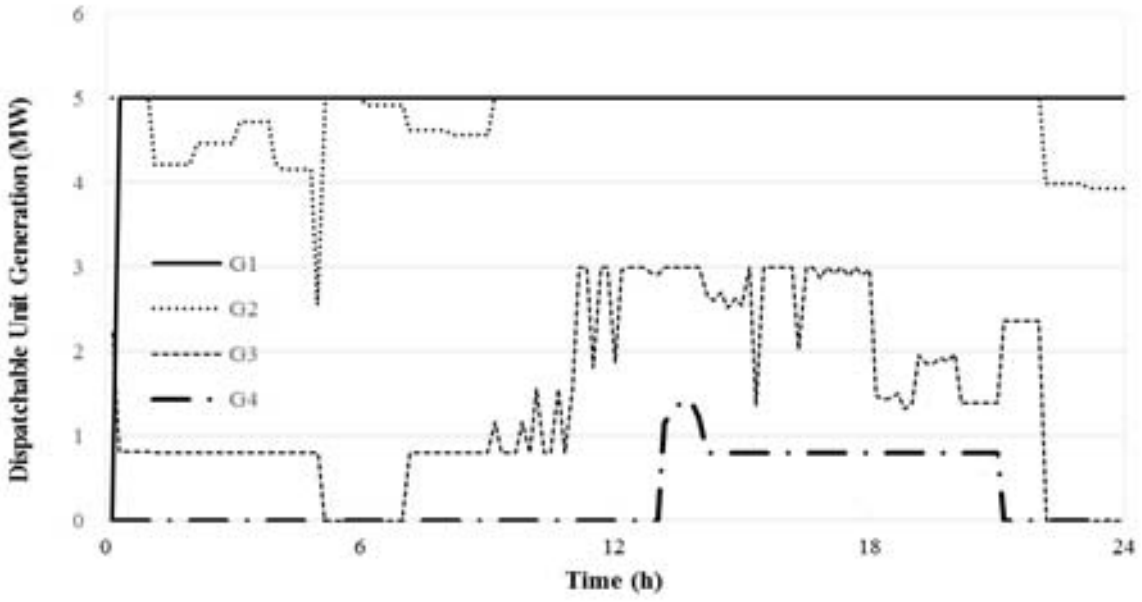

Fig. 4 Generation of different dispatchable units at a scenario with islanding between hours 12-14

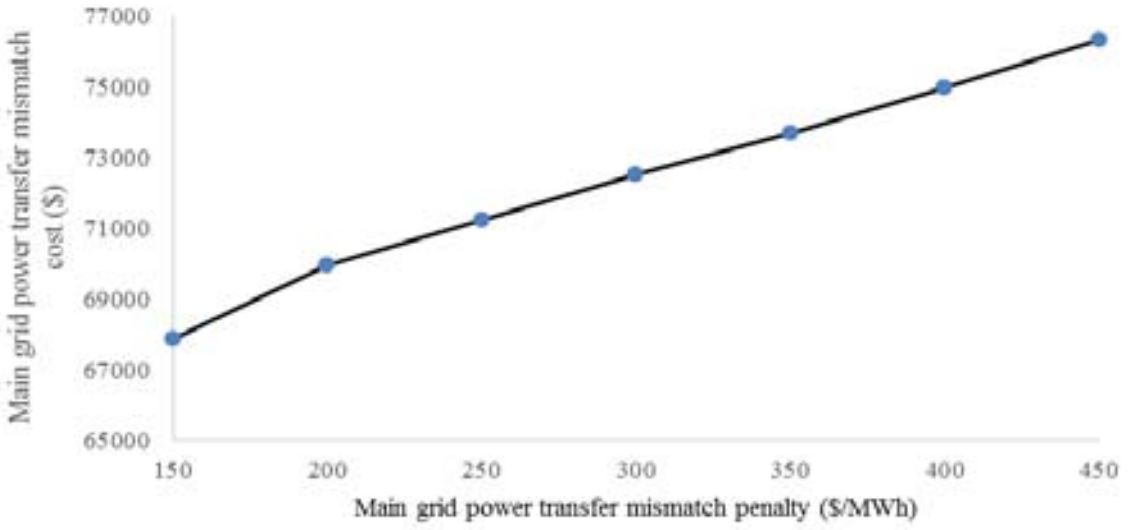

Fig. 5 Microgrid operation cost with different penalties for excess main grid power transfer.

As the power transfer increases, however, previously committed units are turned off at more hours and therefore load curtailment cost for scenarios with islanding could increase. When the main grid power transfer is increased from 1.25 times values in Table I to 1.4, some units are turned off at high load hours 19-21 and 23 leading to load curtailment cost increasing from $4,131 to $11,069 while generation cost reduces from $37,029 to $32,235. With a smaller main grid power transfer, dispatchable units are committed at more hours but generate at the minimum power in the normal operation. Fig. 4 shows the generation of different dispatchable units for a scenario where islanding occurs between hours 12 and 14.

Up to this point, all simulations were conducted for the microgrid with an infinite power transfer mismatch penalty. Enabling the microgrid with the capability to increase its main grid power transfer beyond the amount assigned to it by the DMO and instead paying a penalty would eliminate the load curtailment that otherwise might have been needed. This feature reduces the operation cost significantly. Fig. 5 depicts the increase in operation cost as this penalty increases. It is seen that with lower penalties for transferring additional amounts of main grid power than scheduled, the total operation cost of the microgrid drops.

## V. Conclusion

A market-based microgrid optimal scheduling model was proposed in this paper. Subhourly dispatch was employed to achieve the most economical schedule of microgrid DERs and loads while taking nondispatchable generation variations into account and making sure that the main grid power transfer scheduled by the DMO is achieved. Stochastic optimization was used to account for uncertainties due to islanding and variations in loads and nondispatchable generation. Simulations were performed using CPLEX and the obtained results were studied to show how microgrid can be optimally scheduled while taking distribution market decisions into consideration.